# Fragmentation of CDF jets: perturbative or non-perturbative?

Alexei N. Safonov

For CDF Collaboration

University of Florida

Abstract

Proceedings. International Euroconference in Quantum Chromodynamics 7-13 July 1999, Montpellier, France

# Fragmentation of CDF jets: perturbative or non-perturbative?

Alexei Safonov

Department of Physics, University of Florida,
Gainesville FL32611-8440, USA

For CDF Collaboration

Presented are the most recent jet fragmentation results from CDF: inclusive distributions of charged particle momenta and their $k_T$ in jets; average track multiplicities, as well as angular distributions of multiplicity flow, for a wide range of jet energies with $E_T$ from 40 to 300 GeV. The results are compared with Monte-Carlo and, when possible, analytical calculations performed in resummed perturbative QCD approximations (MLLA).

## 1. INTRODUCTION

Fig. 1 gives a symbolic depiction of the jet fragmentation, in which one can distinguish the stage governed by perturbative QCD and the stage of phenomenological hadronization. The fuzzy boundary between the stages is usually associated with a $k_T$ cut-off scale $Q_0$. To proceed with perturbative calculations, one has to set $Q_0 \gg \Lambda_{QCD}$ so that $\alpha_s$, which is also effectively enlarged by a double log term that comes from soft (dk/k) and collinear ($dk_T/k_T$) divergences, remains small.

**Two Stages of jet evolution:**

**I  - perturbative QCD  ($k_T > Q_0$)**

**II  - phenomenological hadronization**

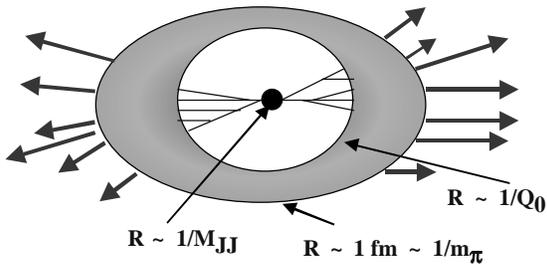

Figure 1. Symbolic depiction of the jet fragmentation.

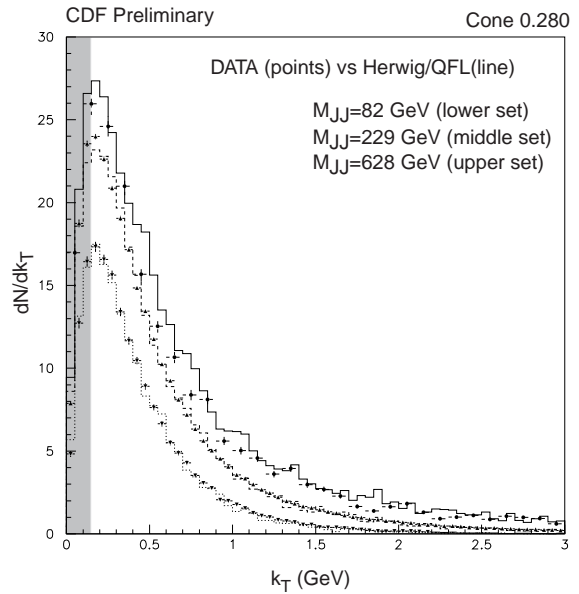

Figure 2. $dN/dk_T$ distribution for tracks within cone 0.28 around the jet axis. Dijet events. Herwig 5.6 scaled by 0.89.

From Fig. 2, one can easily see that within "straightforward" perturbative QCD (cut-off scale of the order of 1 GeV) there is no chance to coherently describe the jet fragmentation, as the vast majority of hadrons within jets has $k_T$ well below 1 GeV scale. Therefore, perturbative methods may seem to be insufficient for predicting the characteristics of jet hadrons and one needs to rely heavily on the hadronization models.

In order to make perturbative methods usable to describe jet fragmentation, two requirements have to be fulfilled: first, final hadrons have to "remember" properties of partons and, second, perturbative calculations have to include partons with sufficiently low $k_T$. In other words, the cut-off scale needs to be pushed down below 1 GeV.

Possible perturbative dominance scenario satisfying the requirements listed above, can be realized by taking perturbative calculations carried out in the framework of Modified Leading Log Approximation (MLLA) [1] together with the hypothesis of Local Parton-Hadron Duality (LPHD) [2].

MLLA calculations (at least, those related to momentum distributions) are infrared stable, which allows one to push the cut-off scale $Q_{eff}$ all the way down to $\Lambda_{QCD}$. In this way, soft partons are accounted for, and the cut-off parameter $Q_{eff}$ has to be obtained from data.

LPHD assumess that hadronization happens locally in the very last moment, so that properties of final hadrons resemble those of initial partons. Thus, hadron distributions are related to parton ones. Parameter *Const*, which basically relates the number of final hadrons and initial partons becomes another free parameter:

$$Const = \frac{N_{hadrons}}{N_{partons}}$$

As a result, we have a perturbative model, which potentially may describe jet fragmentation with only two free parameters - *Const* and $Q_{eff}$.

## 2. THEORETICAL PREDICTIONS

MLLA+LPHD gives a number of very firm predictions, which can be compared to the experimental data.

Here, when comparing data to the MLLA calculations, we will be concentrating mostly on inclusive multiplicity of charged hadrons in jets, the inclusive momentum distribution shape and its peak position.

The MLLA parton momentum spectrum for a gluon jet is given by [1,3]:

$$\frac{1}{\sigma}\frac{d\sigma}{d\xi} = \frac{4Nc}{b}\Gamma(B)\int_{-\pi/2}^{+\pi/2}\frac{d\tau}{\pi}\left[\frac{\cosh\alpha + (1-2\zeta)\sinh\alpha}{\frac{4Nc}{b}Y\frac{\alpha}{\sinh\alpha}}\right]^{B/2} \times$$

$$\times e^{-B\alpha} I_B\left(\sqrt{\frac{4Nc}{b}Y\frac{\alpha}{\sinh\alpha}(\cosh\alpha + (1-2\zeta)\sinh\alpha)}\right)$$

with important variables:

$$\xi = \ln\frac{1}{x}, \quad x = \frac{p}{E_{jet}}, \quad Y = \ln\frac{E_{jet}\theta_{cone}}{Q_{eff}},$$

and some secondary ones:

$$\zeta = 1 - \frac{\xi}{Y}, \quad \alpha = \alpha_o + i\tau, \quad \tanh\alpha_o = 2\zeta - 1$$

It should be pointed out that MLLA calculations are carried out using the assumption $x \ll 1$. On the other side of the spectrum, still one has to stay above p=0.5 GeV to avoid the finite mass effects since MLLA deals with massless partons, not hadrons. Although the curve should go to zero at $\xi=0$, the exact shape of the descending is not controlled in the calculation.

The peak position of the momentum spectrum is given by [3]:

$$\xi_o = \frac{1}{2}Y + \sqrt{cY} - c, \quad \text{where } c = 0.29 \text{ for } n_f = 3$$

The value of the third term in the equation above, the constant, is determined by the MLLA spectrum function. Though, strictly speaking, next-to-MLLA terms may contribute to the peak position at this level.

Multiplicity of partons in a gluon jet in MLLA is given by [3]:

$$N_g = \Gamma(B)\left(\frac{z}{2}\right)^{-B+1} I_{B+1}(z), \quad \text{where } z = \sqrt{16N_cY/b}$$

It is also very important to mention that in MLLA, gluon and quark jets are identical apart from a numerical factor *r*, ratio of multiplicities:

For the range of CDF jet energies, the theoretical value of r is almost constant. In MLLA r=9/4 [3] (ratio of gluon and quark color charges), in the next-to-MLLA calculations r=2.05 [4], and in the next-to-next-to-MLLA order two different values were reported: r=2.0 [5] and r=1.85 [6]. Experimental e+e- data suggests that this ratio r is somewhere between 1.0 and 1.5 [7], and may be energy dependent.

## 3. ANALYSIS AT CDF

The following results are based on approximately 100,000 dijet events taken in 1994-96 at CDF. Selected events were required to have two high $E_T$ well-balanced jets in the central region of the detector. Events with three or four jets were also allowed if the additional jets were sufficiently soft ($E_T$ sum of the third and fourth jets less than 15% of total event $E_T$).

One of the significant advantages of high energy hadron machines, such as Tevatron, is that the range of dijet masses one can look at is rather wide. In this analysis, the masses of the dijet events range between 72 and 740 GeV, and the events were subdivided in 9 sub-samples, as shown in Table 1.

| $072 < M_{jj} < 094$ | $<M_{jj}> = 81.8$ |
|---|---|
| $094 < M_{jj} < 120$ | $<M_{jj}> = 105.1$ |
| $120 < M_{jj} < 154$ | $<M_{jj}> = 140.4$ |
| $154 < M_{jj} < 200$ | $<M_{jj}> = 182.7$ |
| $200 < M_{jj} < 260$ | $<M_{jj}> = 228.9$ |
| $260 < M_{jj} < 340$ | $<M_{jj}> = 292.5$ |
| $340 < M_{jj} < 440$ | $<M_{jj}> = 378.0$ |
| $440 < M_{jj} < 570$ | $<M_{jj}> = 487.9$ |
| $570 < M_{jj} < 740$ | $<M_{jj}> = 628.0$ |

Table 1. Dijet masses (GeV) of the 9 sub-samples.

It is important that MLLA predictions have angular dependence. In other words, one can look at distributions of hadrons within fixed opening angle around the jet axis and compare experimental data with the predictions. In our analysis, we chose 5 such restricted cone-sizes: 0.168, 0.217, 0.280, 0.361 and 0.466. Angle is defined as the angle between the jet direction and the cone side.

Unlike e+e- experiments, in our data we have a mixture of quark and gluon jets, which can not be separated from each other without further investigation. We plan to carry out an analysis similar to this one for γ-jet sample, which is enriched with quark jets and, in this way, to separate those contributions.

In our case, the formulas from the previous chapter have to be modified somewhat to take into account the presence of both kinds of jets:

$$N_{hadrons}(\xi) = Const(N_g(\xi)\varepsilon_g + N_q(\xi)\varepsilon_q),$$

where $N_g$ and $N_q$ are the gluon and quark jet spectra. In MLLA, $N_g(\xi)$ and $N_q(\xi)$ are different by a single factor r, therefore:

$$N_{hadrons}(\xi) = Const(\varepsilon_g + 1/r\varepsilon_q) \cdot N_g, \quad or$$
$$N_{hadrons}(\xi) = K \cdot N_g, \quad K = Const(\varepsilon_g + 1/r\varepsilon_q)$$

Note, that $K$ depends on energy only (via the energy dependence of $\varepsilon_q$ and $\varepsilon_g$ - the quark and gluon jet fractions in the sample).

All data distributions were fitted with MLLA gluon spectrum functions. In this work, we report extracted values of $K$ instead of the true MLLA parameter *Const*.

## 4. RESULTS

Below we present our findings on inclusive multiplicity of charged particles in dijet events (i.e. doubled jet multiplicity), and results of fits of inclusive momentum spectrum of charged particles with MLLA theoretical predictions. Results of the fits are represented in terms of peak position of the distribution, and MLLA parameters $Q_{eff}$ and $K$, we also present the results on multiplicity flow, standard dN/dx fragmentation function and $dN/dk_T$ distributions. Analysis of systematics for these results is yet to be carried out.

### 4.1. Charged Multiplicity

In the next plot, we compare our results for charged particle multiplicity within a restricted cone 0.466 around the jet axis to the predictions of Monte-Carlo (Herwig 5.6 + Detector simulation) and to MLLA predictions corresponding to different values of ratio r. Data errors are completely dominated by correlated systematic uncertainties.

Range of values r goes from r=1, which corresponds to the case when quark and gluon jets are indistinguishable, to r=9/4 (the classic MLLA case). Next-to-MLLA and next-to-next-to-MLLA order predictions range between these extreme cases.

Next-to-next-to-MLLA calculations predict r to be either 1.85 or 2.0, while experimental e+e- results give r in the range of 1.0-1.5. Our data suggests that the quark-gluon jet's multiplicity ratio r, defined as described in chapter 3, is somewhere between 1.4 and 2.

Note that Herwig 5.6 predictions were scaled by a factor 0.89 to match the data. However, given the systematic errors, this difference is not very significant (about 2 sigma).

Throughout our analysis, Herwig 5.6 predictions were always scaled by the same factor 0.89.

Table 2 presents numerical results for multiplicities for all 5 restricted cone-sizes along with both statistical and systematical uncertainties.

One can see that the errors are completely dominated by systematics.

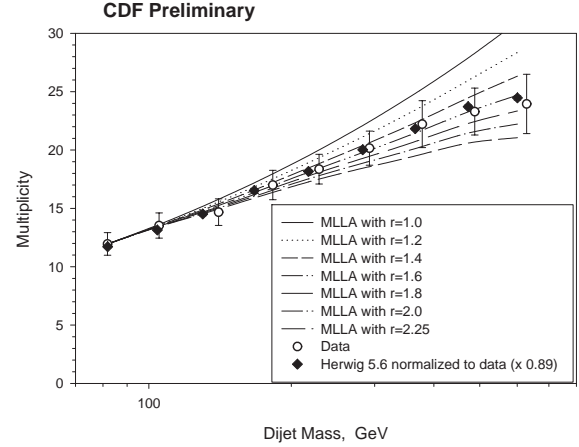

Figure 3. Charged multiplicity within cone 0.466 around the jet axis. Data compared to Monte-Carlo and MLLA predictions with r =1.0 - 2.25.

Fig. 4 allows us to compare the data with Herwig predictions. This plot presents charged multiplicity of particles within 3 restricted cones: 0.168, 0.280, 0.466. One can see that the agreement is very reasonable, apart from the fact that we had to scale Herwig predictions.

CDF Preliminary

| Energy (GeV) | Cone 0.168 | Cone 0.217 | Cone 0.280 | Cone 0.361 | Cone 0.466 |
|---|---|---|---|---|---|
| 81.81 | 5.56±0.06±0.65 | 7.27±0.09±0.73 | 8.85±0.07±0.88 | 10.45±0.08±1.06 | 11.96±0.09±1.27 |
| 105.10 | 6.52±0.09±0.78 | 8.24±0.09±0.84 | 10.08±0.10±1.01 | 11.73±0.11±1.21 | 13.53±0.12±1.43 |
| 140.40 | 7.57±0.08±0.92 | 9.69±0.09±0.98 | 11.47±0.09±1.16 | 13.10±0.09±1.36 | 14.68±0.99±1.57 |
| 182.70 | 8.76±0.08±1.07 | 11.31±0.15±1.14 | 13.27±0.08±1.31 | 15.17±0.09±1.54 | 17.01±0.09±1.80 |
| 228.90 | 10.14±0.07±1.16 | 12.19±0.09±1.24 | 14.36±0.08±1.44 | 16.28±0.08±1.69 | 18.36±0.11±1.97 |
| 292.50 | 11.71±0.10±1.34 | 13.70±0.10±1.39 | 16.06±0.09±1.61 | 18.03±0.10±1.87 | 20.16±0.10±2.16 |
| 378.00 | 13.01±0.12±1.48 | 15.21±0.13±1.53 | 17.56±0.13±1.76 | 20.07±0.14±2.06 | 22.22±0.17±2.37 |
| 487.90 | 14.63±0.17±1.59 | 16.41±0.18±1.66 | 18.56±0.18±1.86 | 20.80±0.19±2.15 | 23.30±0.20±2.49 |
| 628.00 | 15.69±0.29±1.64 | 16.48±0.31±1.68 | 19.10±0.32±1.93 | 21.52±0.35±2.25 | 23.95±0.36±2.60 |

Table 2. Inclusive charged particle multiplicity in dijet events in cones 0.168, 0.217, 0.280, 0.361 and 0.466 with corresponding statistical and systematical errors.

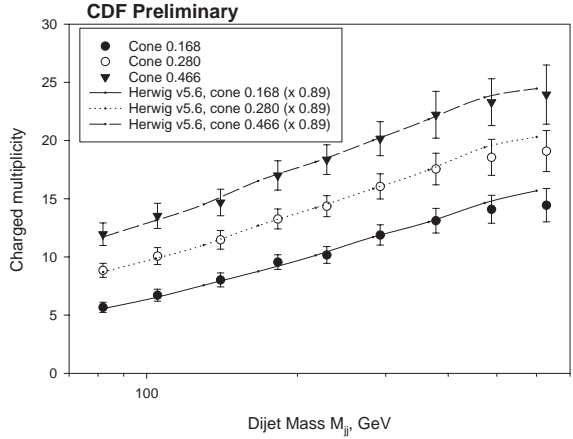

Figure 4. Charged multiplicity of particles within cones 0.168, 0.280, 0.466. Herwig is scaled by a factor 0.89.

### 4.2. Inclusive momentum distribution

Fig.5 presents 9 plots corresponding to 9 dijet mass sub-samples for inclusive momentum distribution of charged particles within cone 0.168 around the jet axis. Data is fitted with the MLLA limiting spectrum function. Fig. 6 presents a similar set of plots but for the largest available cone-size 0.466.

We fitted 45 distributions - all possible combinations - 9 dijet mass sub-samples times 5 cone-sizes. Fitted values for $Q_{eff}$ are presented in Table 3, and the results for the parameter K can be found in Table 4. In Table 5, corresponding fit values of $\chi^2$ are shown.

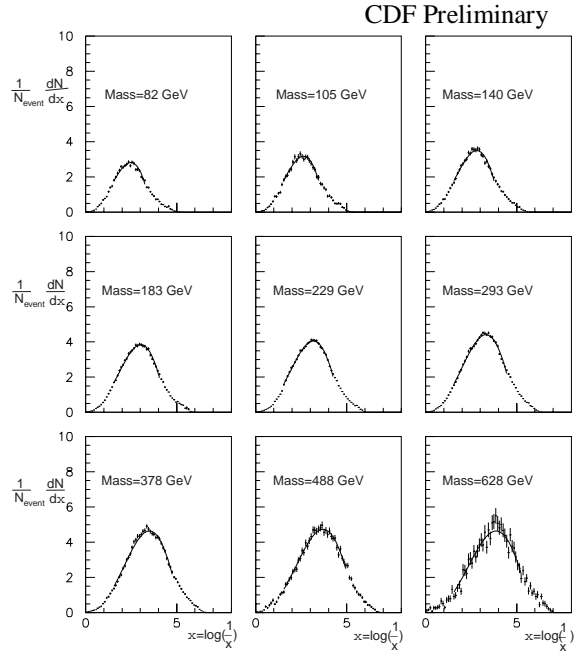

Figure 5. Inclusive momentum distribution of charged particles within cone 0.168 fitted with MLLA function. All available dijet mass samples. Results of the fits are in Tables 3 - 5.

One can clearly see that, for both parameters, the effects of systematics dominate over statistical fluctuations. Though the general agreement between the data and MLLA predictions is rather striking, one can see a clear small excess of data over the MLLA predictions to the left of the peak of the distribution. This is especially visible in the case of high-statistics dijet mass sub-samples.

CDF Preliminary

| Energy (GeV) | Cone 0.168 | Cone 0.217 | Cone 0.280 | Cone 0.361 | Cone 0.466 |
|---|---|---|---|---|---|
| 81.81 | 0.246±0.001±0.039 | 0.263±0.001±0.042 | 0.250±0.005±0.041 | 0.262±0.001±0.045 | 0.280±0.003±0.045 |
| 105.10 | 0.245±0.003±0.039 | 0.221±0.002±0.036 | 0.247±0.001±0.041 | 0.261±0.002±0.044 | 0.279±0.001±0.044 |
| 140.40 | 0.222±0.002±0.035 | 0.252±0.001±0.041 | 0.256±0.001±0.042 | 0.247±0.001±0.042 | 0.265±0.001±0.042 |
| 182.70 | 0.214±0.004±0.034 | 0.224±0.002±0.036 | 0.237±0.003±0.039 | 0.243±0.001±0.041 | 0.260±0.003±0.041 |
| 228.90 | 0.210±0.003±0.033 | 0.220±0.003±0.036 | 0.226±0.001±0.037 | 0.236±0.001±0.040 | 0.251±0.002±0.040 |
| 292.50 | 0.237±0.004±0.038 | 0.227±0.001±0.037 | 0.230±0.003±0.038 | 0.244±0.002±0.041 | 0.250±0.002±0.040 |
| 378.00 | 0.221±0.003±0.035 | 0.222±0.004±0.036 | 0.230±0.003±0.038 | 0.227±0.002±0.039 | 0.245±0.004±0.039 |
| 487.90 | 0.203±0.004±0.032 | 0.200±0.002±0.032 | 0.212±0.004±0.035 | 0.227±0.002±0.039 | 0.241±0.005±0.038 |
| 628.00 | 0.180±0.011±0.029 | 0.191±0.005±0.031 | 0.199±0.009±0.033 | 0.206±0.009±0.035 | 0.227±0.008±0.036 |

Table 3. Fitted values for MLLA parameter $Q_{eff}$ along with statistical and systematical errors.

| Energy (GeV) | Cone 0.168 | Cone 0.217 | Cone 0.280 | Cone 0.361 | Cone 0.466 |
|---|---|---|---|---|---|
| 81.81 | 0.490±0.006±0.060 | 0.557±0.006±0.062 | 0.590±0.007±0.057 | 0.632±0.005±0.061 | 0.661±0.005±0.065 |
| 105.10 | 0.515±0.007±0.063 | 0.528±0.008±0.059 | 0.596±0.007±0.058 | 0.634±0.006±0.062 | 0.659±0.006±0.065 |
| 140.40 | 0.494±0.005±0.060 | 0.549±0.006±0.061 | 0.581±0.005±0.056 | 0.592±0.005±0.058 | 0.611±0.004±0.060 |
| 182.70 | 0.485±0.005±0.059 | 0.529±0.004±0.059 | 0.560±0.004±0.054 | 0.579±0.003±0.056 | 0.598±0.004±0.059 |
| 228.90 | 0.475±0.004±0.058 | 0.510±0.004±0.057 | 0.540±0.003±0.052 | 0.561±0.003±0.055 | 0.584±0.003±0.058 |
| 292.50 | 0.506±0.005±0.062 | 0.520±0.003±0.058 | 0.546±0.004±0.053 | 0.568±0.003±0.055 | 0.575±0.003±0.057 |
| 378.00 | 0.477±0.005±0.058 | 0.503±0.005±0.056 | 0.525±0.008±0.051 | 0.535±0.005±0.052 | 0.551±0.005±0.054 |
| 487.90 | 0.443±0.006±0.054 | 0.463±0.006±0.051 | 0.484±0.006±0.047 | 0.502±0.005±0.049 | 0.518±0.005±0.051 |
| 628.00 | 0.385±0.011±0.047 | 0.411±0.009±0.046 | 0.433±0.010±0.042 | 0.448±0.009±0.044 | 0.468±0.009±0.046 |

Table 4. Fitted values for MLLA related parameter $K$ along with statistical and systematical errors.

As it was already mentioned, the data was fitted as if all jets were gluon jets. Within MLLA, the quark and gluon jet distributions have almost the same shape and differ by only a normalization constant. As a consequence, the fitted parameter $K$, though closely related, is not exactly the MLLA+LPHD constant $Const$.

Therefore, our fits are valid, but the interpretation of the constant $K$ can be done only if the ratio of normalization constants for quark and gluon jets is known.

If quark jets have smaller multiplicity as predicted by MLLA (see discussion of results on multiplicity), then the constant $K$ should fall with jet energy, and it does.

Unlike $K$, parameter $Q_{eff}$ can be analyzed directly, in MLLA $Q_{eff}$ must be a constant. Still, some systematic trends in the behavior of $Q_{eff}$ are visible: it tends to become smaller for larger jet energies.

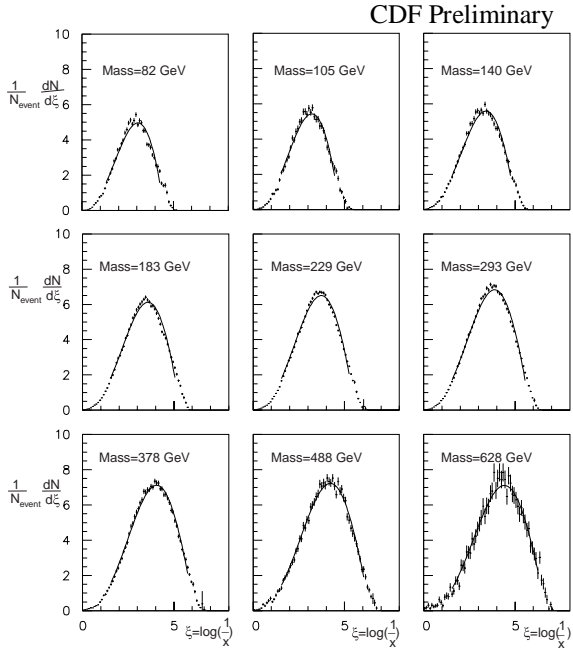

Figure 6. Inclusive momentum distribution of charged particles within cone 0.466 fitted with MLLA function. All available dijet mass samples.

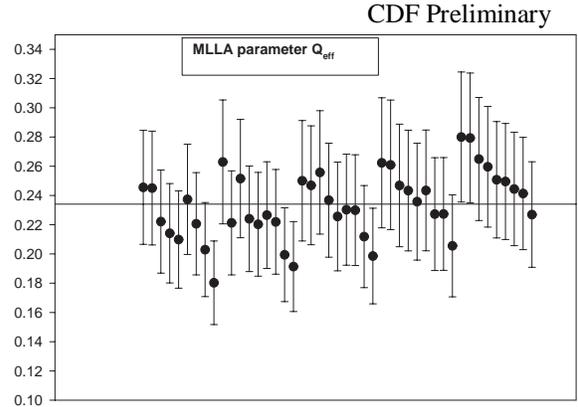

Figure 7. Fitted values for $Q_{eff}$ corresponding to 45 (5 cones x 9 dijet mass cuts) possible combinations. Five series of data points correspond to five cone sizes, 9 data points within each cone size series correspond to different dijet masses (lowest at left). $Q_{eff}$ = 240 ±40 MeV.

As far as parameter *K* concerned, it was shown that *K* depends on energy only. Thus, for the same dijet mass sample, *K* should stay constant for all cone-sizes.

Data in Table 4 indicates that this is violated. The plot presented on Fig. 8 makes this observation apparent. We plotted and fitted momentum distributions, corresponding to three different cone-sizes, corresponding to the same dijet mass $M_{JJ}$=378 GeV.

However, the assumption that r is energy independent may be incorrect as indicated by LEP experiments. In this case one needs to do a more thorough analysis that would include the γ(Z/W)-jet sample. This will allow for direct measurement of the ratio r and unambiguous extraction of the MLLA *Const* from the fit parameter *K*.

### 4.3. Peak position of the inclusive momentum distribution

Fig. 9 shows the peak position of the inclusive momentum distributions plotted against $M_{jj}\sin\theta$ (θ - opening angle).

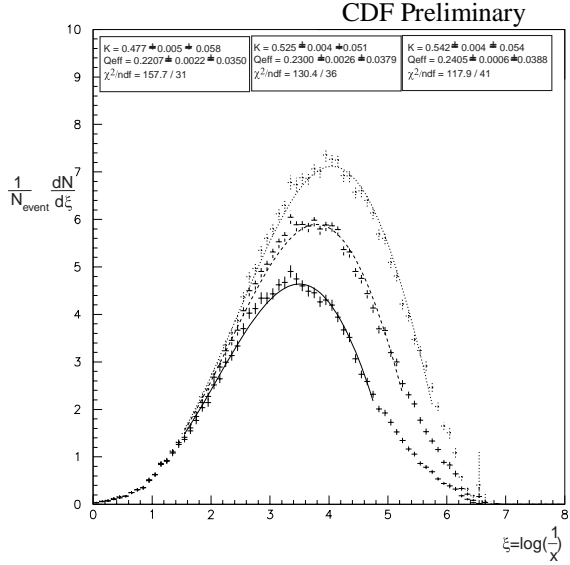

Figure 8. Inclusive momentum distribution of charged particles within cones 0.168, 0.280 and 0.468 fitted with MLLA function. Dijet mass 378 GeV.

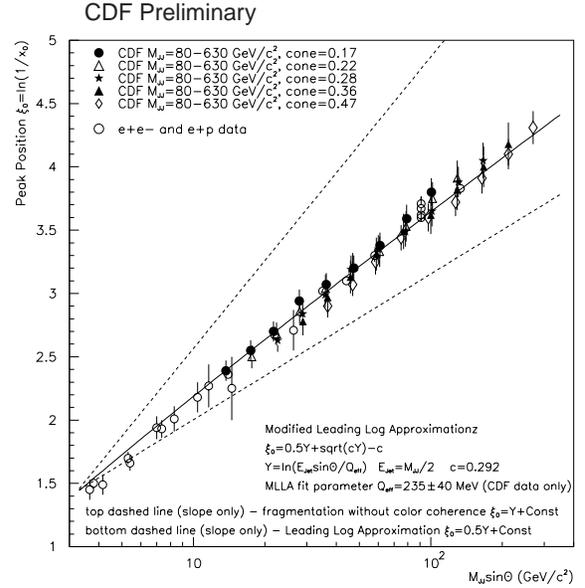

Figure 9. Peak position of the inclusive momentum distributions plotted against $M_{jj}\sin\theta$.

| Energy (GeV) | $\chi^2$ | | | | |
|---|---|---|---|---|---|
| | Cone 0.168 | Cone 0.217 | Cone 0.280 | Cone 0.361 | Cone 0.466 |
| 81.81 | 37.27 / 16 | 50.86 / 18 | 84.58 / 21 | 85.29 / 23 | 128.1 / 26 |
| 105.10 | 19.49 / 18 | 69.31 / 23 | 41.61 / 23 | 47.23 / 26 | 56.63 / 28 |
| 140.40 | 86.03 / 23 | 98.97 / 24 | 94.49 / 26 | 101.4 / 29 | 85.85 / 31 |
| 182.70 | 112.5 / 24 | 150.6 / 26 | 219.7 / 29 | 228.7 / 31 | 272.2 / 34 |
| 228.90 | 48.62 / 26 | 91.56 / 28 | 174.8 / 31 | 257.8 / 34 | 230.6 / 36 |
| 292.50 | 184.1 / 28 | 153.5 / 31 | 227.1 / 33 | 182.1 / 36 | 268.8 / 39 |
| 378.00 | 157.7 / 31 | 117.0 / 34 | 130.4 / 36 | 116.5 / 39 | 117.9 / 41 |
| 487.90 | 57.99 / 33 | 59.89 / 36 | 64.92 / 39 | 73.39 / 41 | 74.14 / 44 |
| 628.00 | 74.53 / 36 | 76.82 / 39 | 82.20 / 41 | 89.31 / 44 | 82.86 / 46 |

Table 5. Account of values of $\chi^2$ obtained for the fits of the inclusive momentum distributions.

The MLLA fit gives $Q_{eff} = 235 \pm 40$ MeV. The error is completely dominated by systematic errors in determination of the peak position. This remarkable agreement with the theoretical expectations in the framework of MLLA is also supported by e+e- and ep data (empty circles). For reference, The Leading Log prediction is also plotted on the same graph, as well as the prediction in the case of fragmentation without color coherence effects

### 4.4. dN/dlogp distribution

When plotted as dN/dln(p), Fig. 10, the MLLA inclusive momentum distributions reveal an interesting feature: the multiplicity of soft tracks in a jet is almost jet energy independent. This is a direct consequence of the color interference. To understand this, first it is important to remember that the probability of a soft gluon emission does not depend directly on the jet energy. Second, one needs to recall that the color interference can be effectively accounted for as an angular ordering of emissions: emissions should occur at subsequently decreasing angles.

So, even if all emissions occur at the minimum $k_T$ (it is natural to assume that the minimum $k_T$ is of the order of $Q_{eff} = \Lambda_{QCD}$), then all further emitted gluons would have to have larger and larger energies. Eventually, such emissions are terminated as the result of phase space limitations and energy conservation. The larger the jet energy is, the farther away those limits are.

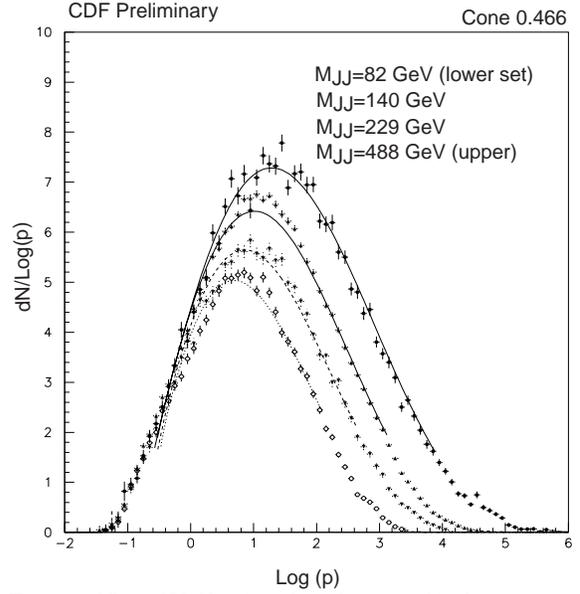

Figure 10. d$N$/d$log(p)$ distribution (4 dijet mass samples, cone 0.466).

Thus, as jet energy increases, the multiplicity grows mostly due to partons with increasingly higher momentum flying out at decreasingly smaller angles.

Fig. 10 shows the dN/dln(p) distribution for the data. One can see that the low momentum region of the spectrum (p<1 GeV) is indeed almost energy independent.

CDF Preliminary

| Energy (GeV) | Cone 0.168 | Cone 0.217 | Cone 0.280 | Cone 0.361 | Cone 0.466 |
|---|---|---|---|---|---|
| 81.81 | 2.39±0.02±0.07 | 2.50±0.01±0.09 | 2.63±0.02±0.09 | 2.78±0.01±0.11 | 2.90±0.01±0.09 |
| 105.10 | 2.55±0.02±0.08 | 2.68±0.01±0.09 | 2.84±0.02±0.09 | 2.97±0.02±0.12 | 3.07±0.01±0.09 |
| 140.40 | 2.70±0.01±0.08 | 2.86±0.01±0.10 | 2.99±0.01±0.10 | 3.13±0.01±0.13 | 3.25±0.01±0.10 |
| 182.70 | 2.94±0.01±0.09 | 3.04±0.01±0.11 | 3.19±0.01±0.11 | 3.31±0.01±0.13 | 3.44±0.01±0.10 |
| 228.90 | 3.07±0.01±0.09 | 3.21±0.01±0.11 | 3.35±0.01±0.11 | 3.49±0.01±0.14 | 3.60±0.01±0.11 |
| 292.50 | 3.20±0.01±0.10 | 3.33±0.01±0.12 | 3.49±0.01±0.12 | 3.62±0.01±0.15 | 3.72±0.01±0.11 |
| 378.00 | 3.38±0.01±0.10 | 3.53±0.01±0.12 | 3.65±0.01±0.12 | 3.82±0.02±0.15 | 3.91±0.01±0.12 |
| 487.90 | 3.59±0.02±0.11 | 3.75±0.02±0.13 | 3.88±0.02±0.13 | 4.00±0.02±0.16 | 4.10±0.02±0.12 |
| 628.00 | 3.80±0.03±0.12 | 3.91±0.03±0.14 | 4.05±0.04±0.13 | 4.18±0.04±0.17 | 4.31±0.03±0.13 |

Table 6. Account of systematic errors for Peak Position of the inclusive momentum distribution.

## 4.5. Multiplicity flow

Fig. 11 shows the differential $dN/d\theta$ distribution of charged particles in the jet compared to Monte-Carlo predictions (Herwig 5.6 along with the detector simulation). One can see that agreement is very reasonable.

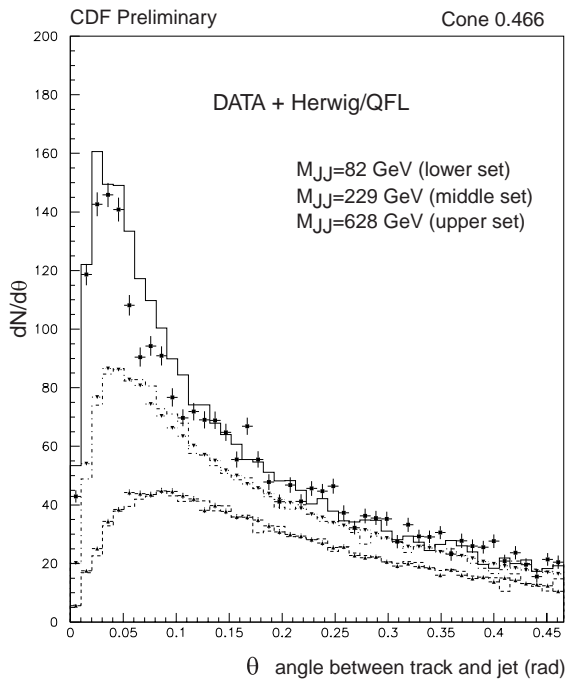

Figure 11. $dN/d\theta$ distribution of charged particles in dijets compared to Monte-Carlo (Herwig 5.6 scaled by 0.89 along with the detector simulation).

## 4.6. dN/dx distribution

So far we have been analyzing the jet fragmentation with the emphasis on the soft momentum tracks. Fig. 12 shows the standard $dN/dx_p$ fragmentation function for three different di-jet masses (tracks are from the cone $\theta_{cone}=0.47$).

It is interesting to compare these results to the e+e- fragmentation functions. Fig. 13 shows the CDF $M_{jj}$=82 GeV data (same data points as in Fig. 12) plotted together with TASSO [8] and DELPHI [9] data points.

One should not compare the curves on Fig. 13 at very low $x_p$ - CDF data are from restricted cones around the jet axis and, therefore, are bound to lack soft tracks compared to the e+e- data, where tracks from the full solid angle are included in the plots.

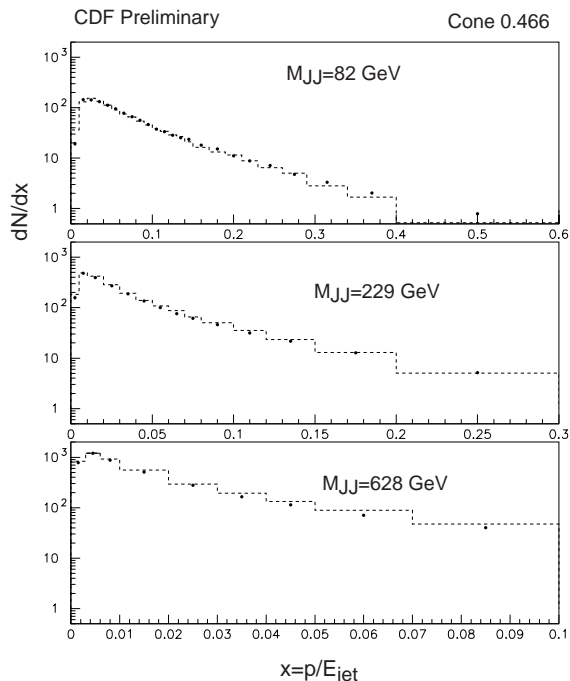

Figure 12. Standard $dN/dx$ distribution compared to Monte-Carlo predictions.

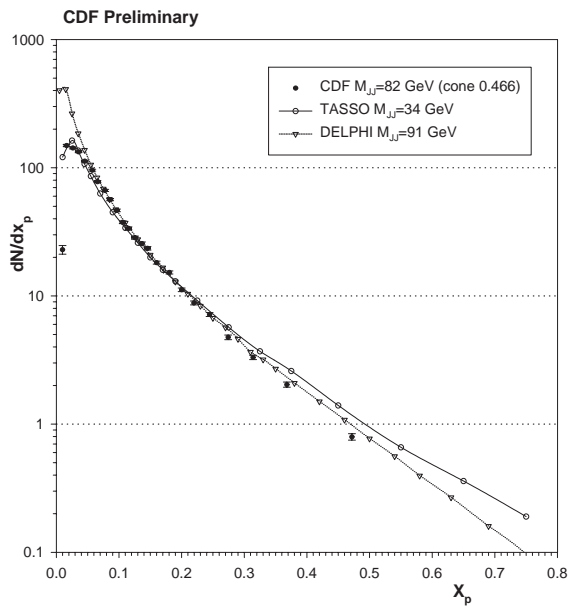

Figure 13. Fragmentation function $dN/dx$. Comparison of the CDF data to TASSO [8] and DELPHI [9] results.

On the other hand, the points at larger x can be compared (the high momentum tracks mostly come from the very collimated region around the jet axis). From the QCD scaling violation, one expects that at the domain of larger x, the curves should go lower as the jet energy increases (cf. TASSO and DELPHI results). However, one can see that at larger x, CDF data points tend to go lower than the DELPHI points, although the CDF dijet data sample plotted here has slightly lower energy.

The reason for this observation may be twofold. First, the CDF jets at Mjj=82 GeV are predominantly (~75%) gluon jets, while the e+e- jets are quark jets. As was discussed above, the gluon jets should yield higher multiplicity in the low-x region. From energy conservation, this immediately implies that the high-x region should be somewhat depleted. Second, the e+e- analysis is done for all tracks from the full solid angle, which will include all multi-jet events. Those will contribute a certain number of hard tracks.

Cross-comparison of γ-jet and jet-jet events will shed more light on differences in quark and gluon jets. Meanwhile, Fig. 12 shows that the data and Herwig/QFL are in very good agreement (apart from the overall normalization 0.89 used for Monte-Carlo).

## 5. CONCLUSIONS

Our findings can be briefly summarized as follows:

- MLLA gives reasonable agreement with data (multiplicity N, dN/dξ distribution and the peak position evolution). For the cut-off parameter, we report value $Q_{eff}$=240±40 MeV.
- Indirectly measured gluon/quark jet multiplicity ratio r, defined assuming that quark and gluon jets have the same shape (as in MLLA), seems to be between 1.4 and 2.0. Future jet-jet vs γ-jet analysis will allow for direct measurement of r, and is in progress.
- Herwig 5.6 was found to be in a good agreement with data (apart from a scaling factor 0.89, which is not very significant given the systematic uncertainties) over the wide range of analyzed distributions of hadrons in jets.

## REFERENCES

[1] Yu. Dokshitzer, S. Troyan, XIX Winter School of LNPI, vol. 1, p. 144; A. Mueller, Nucl. Phys. B213 (1983) 85; Erratum quoted ibid., B241 (1984) 141.

[2] Ya. I. Azimov, Yu. Dokshitzer, V. Khoze, and S.troyan, Z. Phys. C27 (1985) 65 and C31 (1986) 213.

[3] Yu. Dokshitzer, V. Khoze, A.H. Mueller, S. Troyan, Basics of Perturbative QCD, 1991, Editions Frontiers, France

[4] A.H. Mueller, Nucl. Phys. B241 (1984),141; E.D. Malaza, Z. Phys. C31 (1986), 143

[5] J. B. Gaffney and A.H. Mueller, Nucl. Phys. B250 (1985) 109

[6] I.M. Dremin, V.A. Nechitailo, Modern Phys. Letters A, vol. 9, No. 16 (1994) 147

[7] OPAL Collaboration, CERN-PPE/95-075

[8] TASSO Collaboration, Z. Phys. C47 (1990) 187, Z.Phys. C22 (1984) 307

[9] DELPHI Collab., Phys. Lett. B275 (1992) 231, Z.Phys. C70 (1996) 179, Physics Lett. B311 (1993) 408

## QUESTIONS

*R. Peschanski, Saclay:*

To which extent there is a correlation between using MLLA for the perturbative calculation and the determination of the quark/gluon multiplicity ratio r?

*A. Safonov:*

Our definition of r heavily relies on MLLA calculations. Basic assumption used was that quark and gluon inclusive momentum spectra have the same shape and only differ by a normalization factor.